\documentclass[12pt]{iopart}

\usepackage{graphicx}
\usepackage{color}
\usepackage{soul}
\usepackage{url}
\usepackage{bm}         
\usepackage{times}
\usepackage{dcolumn}
\usepackage{bm}
\usepackage{epsf}
\usepackage{amssymb}

\newcommand{\beq}{\begin{equation}}
\newcommand{\eeq}{\end{equation}}
\newcommand{\beqn}{\begin{eqnarray}}
\newcommand{\eeqn}{\end{eqnarray}}

\usepackage{color}

\begin{document}

\title[Precessing NSBH ejecta]{Dynamical ejecta from precessing neutron star-black hole mergers with a hot, nuclear-theory based equation of state}

\author{
F. Foucart$^{1,2}$,
D. Desai$^3$,
W. Brege$^4$,
M.D. Duez$^4$,
D. Kasen$^{1,3,5}$,
D.A. Hemberger$^6$,
L. E. Kidder$^7$,
H.P. Pfeiffer$^8$,
M.A. Scheel$^6$
}

\address{$^1$ Nuclear Science Division, Lawrence Berkeley National Laboratory, 1 Cyclotron Rd, Berkeley, CA 94720, USA}
\address{$^2$ NASA Einstein Fellow}
\address{$^3$ Department of Physics, University of California, Berkeley, Le Conte Hall, Berkeley, CA 94720}
\address{$^4$ Department of Physics \& Astronomy,	Washington State University, Pullman, Washington 99164, USA}
\address{$^5$ Astronomy Department and Theoretical Astrophysics Center, University of California, Berkeley, 601 Campbell Hall, Berkeley CA, 94720}
\address{$^6$ TAPIR, Walter Burke Institute for Theoretical Physics, MC 350-17, California Institute of Technology, Pasadena, California 91125, USA}
\address{$^7$ Center for Radiophysics and Space Research, Cornell University, Ithaca, New York, 14853, USA}
\address{$^8$ Canadian Institute for Theoretical Astrophysics, University of Toronto, Toronto, Ontario M5S 3H8, Canada}

\begin{abstract}
Neutron star-black hole binaries are among the strongest sources of gravitational waves detectable by current observatories. They can
also power bright electromagnetic signals (gamma-ray bursts, kilonovae), and may be a significant source of production of r-process nuclei. 
A misalignment of the black hole spin with respect to the orbital angular momentum leads to precession of that spin and of the orbital
plane, and has a significant effect on the properties of the post-merger remnant and of the material ejected by the merger. We present
a first set of simulations of precessing neutron star-black hole mergers using a hot, composition dependent, nuclear-theory based equation of state (DD2). 
We show that the mass of the remnant
and of the dynamical ejecta are broadly consistent with the result of simulations using simpler equations of state, 
while differences arise when considering the dynamics of the merger and 
the velocity of the ejecta. We show that the latter can easily be understood from assumptions about the composition of low-density, cold
material in the different equations of state, and propose an updated estimate for the ejecta velocity which takes those effects into account.
We also present an updated
mesh-refinement algorithm which allows us to improve the numerical resolution used to evolve neutron star-black hole mergers. 
\end{abstract}

\pacs{04.25.dg, 04.40.Dg, 26.30.Hj, 98.70.-f}

\maketitle

\section{Introduction}
\label{sec:intro}

The merger of a neutron star-black hole (NSBH) binary is one of the strongest source of gravitational waves detectable
by the current generation of ground-based observatories (Advanced LIGO~\cite{aLIGO2}, Advanced Virgo~\cite{aVirgo2}, KAGRA~\cite{kagra}). 
The recent detection
by advanced LIGO of two merging black hole binaries opened the era of gravitational wave astronomy~\cite{LIGOVirgo2016a,Abbott:2016nmj}. 
With a growing number
of detectors and the improved sensitivities expected in the coming years, many more binary black holes will probably be observed~\cite{2016arXiv160604856T}. 
Gravitational waves from mergers involving neutron stars (binary neutron stars and/or NSBH binaries) are also expected
in the near future~\cite{Abadie:2010cfa}.

In the presence of at least one neutron star, information complementing the gravitational wave signal
can be obtained from powerful electromagnetic signals
which are expected to follow the mergers. Short gamma-ray bursts are generally thought to be powered by neutron
star mergers~\cite{moch:93,LK:98,Janka1999,Fong2013}, 
while the ejection of neutron rich material by some mergers can produce bright optical/infrared transients
on a timescale of days after the merger (a {\it kilonova}, or {\it macronova})~\cite{1976ApJ...210..549L,Li:1998bw,Roberts2011,Kasen:2013xka,Tanaka:2013ana}. 
These electromagnetic signals could provide information about
the localization and environment of the merger, as well as the parameters of the merging objects.
Additionally, r-process nucleosynthesis in the ejected material results in the production of many heavy elements whose
origin remains uncertain today~\cite{korobkin:12,Wanajo2014}. 
Determining whether neutron star mergers are the main site of production of r-process elements
is an important open problem in nuclear astrophysics today. The observation of neutron star mergers can also help us constrain
the equation of state of matter in the cold, dense, neutron rich core of neutron stars~\cite{Read2009b,DelPozzo:13,Lackey2014,Clark:2015zxa,Takami:2015}, 
and thus probe the strength of nuclear forces in
conditions otherwise inaccessible to us.

General relativistic simulations of compact binary mergers are critical to understand the dynamics of mergers, and the
properties of the post-merger remnant and ejected material. They are also needed to obtain reliable templates for the 
gravitational wave signals emitted by mergers. Numerical simulations so far have shown that for circular binaries,
a NSBH merger can result either in the rapid disruption of the neutron star by the black hole, or a direct plunge
of the neutron star into the black hole as the neutron star reaches the innermost stable circular orbit (ISCO) of the black hole.
The competition between the binary separation at tidal disruption and the location of the ISCO is thus the main determinant
of the outcome of the merger. Practically, the most important parameters in the determination of the merger
outcome are the radius of the neutron star (larger stars favor disruption), the mass of the black hole (more massive
black holes hinder disruption), and the spin of the black hole (a larger aligned component of the black hole
spin helps disruption)~\cite{Pannarale:2010vs,Foucart2012}. 

If the neutron star is tidally disrupted, some material is immediately unbound by the
merger ({\it dynamical ejecta})~\cite{Foucart:2013a,kyutoku:2015}, 
some material is placed on highly eccentric bound orbits ({\it bound tail}), some forms a hot
accretion disk around the black hole, and most of the matter is accreted by the black hole within $\sim 1\,{\rm ms}$.
Long term evolutions of remnant accretion disks show that magnetically-driven winds~\cite{Kiuchi:2015qua}, 
neutrino-driven winds~\cite{Dessart2009,Just2014}, 
and energy deposition due to magnetic turbulence and nuclear recombination~\cite{Fernandez2013,Just2014} can combine to unbind a significant
fraction of the remnant disk over timescales of up to a few seconds ({\it disk outflows}, for $\sim 5-20\,\%$ of the disk mass at least). 
One of the main objective of merger simulations is to determine the mass and properties (velocity, composition, entropy)
of the dynamical ejecta and disk outflows, as these have a significant impact on the brightness, color, and duration
of kilonovae~\cite{2013ApJ...775...18B,Barnes:2016}, and on the abundance of the various elements produced through r-process 
nucleosynthesis~\cite{Wanajo2014,Lippuner2015}.

Our current ability to accurately predict the result of a NSBH merger is largely limited by two factors: the complexity of the physical processes
which should be included in numerical simulations, and the wide parameter space that we need
to explore. The most important parameters to vary are the masses of the compact objects,
the spin magnitude and orientation of the black hole, and the unknown equation of state of the neutron star~\cite{Foucart2012}. Neutron star
spins and the orbital eccentricity of the binary~\cite{East:2015yea} could also play an important role, especially for binaries formed through dynamical
interactions in dense clusters. 

The impact of the orientation of the black hole spin has only been studied by general relativistic simulations using
either an ideal gas equation of state~\cite{Foucart:2010eq,Foucart:2013a}, or piecewise polytropic equations of state fitted to nuclear theory for
cold matter in beta-equilibrium, but without any composition information and with a very idealized temperature dependence of the
equation of state~\cite{Kawaguchi2015}. While this provides us with good first estimates of the result of the merger, important properties
of the post-merger remnant are either inaccurate (e.g. velocity and geometry of the dynamical ejecta), or inaccessible 
(e.g. composition, neutrino cooling) without the use of a more advanced description of the neutron star matter. 
In general, at fixed spin magnitude, the main consequence of a misaligned black hole spin is to make the disruption of the neutron star less likely.
This has been observed both in fully general relativistic simulations~\cite{Foucart:2010eq,Foucart:2013a,Kawaguchi2015}, as well as in
smoothed particle hydrodynamics simulations of higher mass ratio binaries performed in a fixed Kerr metric~\cite{Rantsiou:2007ct}.

Here, we study the properties
of the dynamical ejecta and the formation of an accretion disk in a small number of NSBH mergers with misaligned spins using
the hot, composition dependent {\it DD2} equation of state~\cite{Hempel:2011mk}. We can then compare our results with the larger database
of precessing NSBH simulations recently published by Kawaguchi et al.~\cite{Kawaguchi:2015} , as well as the semi-analytical model
for the properties of the ejected matter derived from these simulations~\cite{Kawaguchi:2016}. 

The post-merger evolution of NSBH binaries is significantly affected by both magnetic fields and neutrinos. In recent years,
numerical simulations have rapidly improved their treatment of these complex physical effects in NSBH mergers.
Magnetic fields can power magnetically dominated outflows (``proto-jets'')~\cite{Paschalidis2014},
while neutrinos play an important role in the cooling of the post-merger disk and the evolution of the composition of the fluid~\cite{FoucartM1:2015}.
Both magnetic fields~\cite{Kiuchi:2015qua} and neutrino 
absorption~\cite{Dessart2009,Just2014,FoucartM1:2015} can drive disk winds after the merger. The dynamics of the merger
and the properties of the dynamical ejecta are, on the other hand,
largely unaffected by magnetic fields. The dynamical ejecta is also generally too cold, and moving away from the remnant
too fast, to undergo significant changes of composition due to neutrino-matter interactions. 

For this study of the merger dynamics and tidal disruption of NSBH binaries, we thus choose
to neglect the effects of magnetic fields. Additionally, we only include neutrino cooling through the use of a simple leakage 
scheme~\cite{Ruffert1996,Rosswog:2003rv,OConnor2010},
in which the energy and lepton number carried away by neutrinos is estimated from the local properties of the matter and an
estimate of the optical depth of the fluid. While more advanced neutrino transport schemes have recently been implemented
in merger simulations~\cite{FoucartM1:2015,FoucartM1:2016,Sekiguchi:2015,Sekiguchi:2016}, neutrino-matter interactions will not play any significant role in a 
NSBH merger before the formation of a hot
accretion disk. We only include the neutrino leakage scheme for its ability to approximately capture neutrino cooling in the post-merger
remnant, and because the computational cost of the leakage scheme is small. Details of our implementation of the leakage scheme can
be found in previous publications~\cite{Deaton2013,Foucart:2014nda}.

We first provide a description of our initial conditions in Sec.~\ref{sec:IC}. 
The physical results (merger dynamics, ejecta properties, 
post-merger evolution) are discussed in Sec.~\ref{sec:results}, while Sec.~\ref{sec:discussion} provides a more in-depth comparison with previous results. 
A brief summary and concluding remarks are provided in Sec.~\ref{sec:conclusion}.
We also provide in the Appendix a description of an improved method to adaptively modify our computational grid, and a discussion of the expected numerical 
accuracy of our simulations. 

\section{Initial Conditions}
\label{sec:IC}

\begin{table}
\begin{center}
\caption{Initial parameters of the binaries studied in this paper. $M_{\rm BH}$ is the Christodoulou mass of the black hole,
$M_{\rm NS}$ the ADM mass of an isolated neutron star with the same equation of state and baryon mass as the neutron star under
consideration, $\chi_{\rm BH}$ is the dimensionless spin of the black hole, $i_{\rm BH}$ is the initial inclination between the black hole spin
and the orbital angular momentum, $N_{\rm orbits}$ is the number of orbits up to the point at which $0.01M_\odot$ has been accreted by the black hole,
$\Omega_0$ is the initial angular velocity, and $M=M_{\rm BH}+M_{\rm NS}$. $\Delta x_{\rm dis}$ is the typical grid resolution in the laboratory frame
for the finest level of refinement used during the disruption of the neutron star (see~\ref{sec:grid} for more detail on the grid structure).}
{
\begin{tabular}{c|ccccccc}
Model & $M_{\rm BH}\,(M_\odot)$ & $M_{\rm NS}\,(M_\odot)$ & $\chi_{\rm BH}$ & $i_{\rm BH}$ (deg) & $N_{\rm orbits}$ & $\Omega_0 M$ & $\Delta x_{\rm dis}$ (m) \\
\mr
M5-S7-I60 & 5 & 1.4 & 0.7 & 60 & 3.3 & 0.0416 & 170\\
M5-S9-I60 & 5 & 1.4 & 0.9 & 60 & 3.7 & 0.0415 & 160 \\
M7-S7-I60 & 7 & 1.4 & 0.7 & 60 & 3.1 & 0.0479 & 190\\
M7-S9-I60 & 7 & 1.4 & 0.9 & 60 & 4.2 & 0.0478 & 170\\
M7-S9-I20 & 7 & 1.4 & 0.9 & 20 & 4.8 & 0.0474 & 160\\
\end{tabular}
\label{tab:ID}
}
\end{center}
\end{table}

We consider a series of NSBH binaries in which we modify the black hole mass and spin. We fix the ADM mass of the neutron star (at
infinite separation) to $M_{\rm NS}=1.4M_\odot$,
and use the DD2 equation of state~\cite{Hempel:2011mk}. 
This leads to neutron stars of areal radius $R_{\rm NS}=13.2\,{\rm km}$, at the high end of the range of values
compatible with nuclear theory and astrophysical constraints~\cite{2013ApJ...765L...5S}. The Christodoulou mass of the black hole is either $M_{\rm BH}=5M_\odot$ or 
$M_{\rm BH}=7M_\odot$,
covering the lower range of black hole masses observed in X-ray binaries~\cite{Ozel:2010,Kreidberg:2012}. 
Higher mass black holes do not lead to the disruption
of the neutron star, except for near-extremal, aligned black hole spins. The dimensionless spin of the black hole is either $\chi_{\rm BH}=0.7$ or $\chi_{\rm BH}=0.9$.
The spin of black holes in NSBH binaries is entirely unknown, and the choice of moderate to high spins is made in order to cover the expected
range of spins over which the neutron star disrupts. Finally, the inclination of the spin with respect to the initial angular momentum of the system,
in the coordinates of our initial data,
is set to $i_{\rm BH}=60^{\circ}$, except for a high-mass, high-spin case in which we set $i_{\rm BH}=20^\circ$. A summary of the simulations
considered here is provided in Table~\ref{tab:ID}. In the rest of the text, we label each configuration by the mass of the black hole, the spin of the black hole,
and the inclination of that spin, e.g. M5-S7-I60 corresponds to a simulation with $M_{\rm BH}=5M_\odot$, $\chi_{\rm BH}=0.7$, and $i_{\rm BH}=60^\circ$.

The initial data is generated using our in-house elliptic solver~\cite{Pfeiffer2003,FoucartEtAl:2008}, which includes an iterative algorithm to solve for the constraints 
in Einstein's equations, while requiring that the binary components have the requested masses and spins, that the binary is in a quasi-circular orbit (effectively 
leading to small eccentricities $e\sim 0.01$), and that the fluid is in hydrostatic equilibrium and the flow of the neutron star matter is 
irrotational~\cite{GourgoulhonEtAl2001a,TaniguchiEtAl:2006}. We consider relatively close
initial separations, as we are more interested in the impact of a hot, nuclear theory based equation of state on the dynamics of the merger than in the production
of long, accurate waveforms. The systems considered here evolve for $3-5$ orbits before the neutron star is disrupted by the tidal field of the black hole (or plunges
into the black hole).

\section{Results}
\label{sec:results}

We evolve the coupled system of Einstein's equation of general relativity and the general relativistic equations of hydrodynamics using
the SpEC code~\cite{SpECwebsite}. SpEC uses pseudospectral methods to evolve Einstein's equations in the Generalized Harmonics 
formalism~\cite{Lindblom:2007,Szilagyi:2014fna},
and finite volume methods to evolve the fluid using high-order shock capturing methods. The two systems of equations are coupled
through interpolation of the metric and fluid variables at the end of each time step. The magnitude and orientation of the spin in the simulations
are measured by using approximate rotational Killing vectors~\cite{Lovelace2008}.
A detailed description of our numerical methods can be found in previous publications~\cite{Duez:2008rb,Foucart:2013a}.

\subsection{Late inspiral and merger dynamics}
\label{sec:merger}

\begin{figure}
\begin{center}
\includegraphics*[width=.9\textwidth]{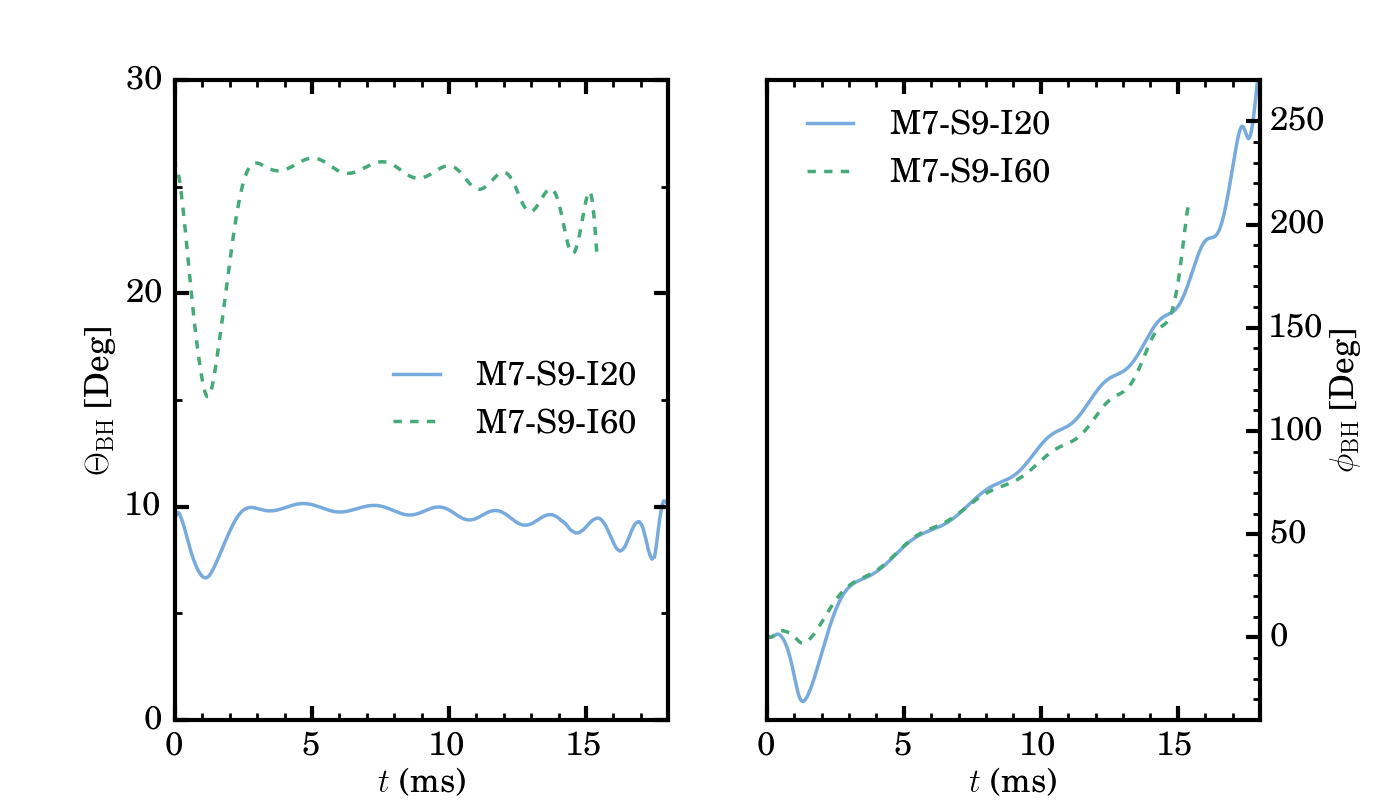}
\caption{Inclination angle $\Theta_{\rm BH}$ of the black hole spin with respect to the initial direction of the total angular momentum of the system ({\it left}), and phase 
$\phi_{\rm BH}$ of the precession of the black hole spin around the total angular momentum ({\it right}). As expected, $\Theta_{\rm BH}$ is nearly constant in time (up to
the impact of the loss of angular momentum due to gravitational wave emission). We show results for the two longest 
simulations, M7-S9-I60 and M7-S9-I20. Over the course of
the evolution, the binary goes through $0.5-0.75$ spin precession periods. The rapid variation of the measured inclination
at early times is due to a smooth change in the gauge
condition between the initial data and the evolution, and shows how coordinate-dependent these measurements can be in a truly dynamical spacetime.}
\label{fig:prec}
\end{center}
\end{figure}

The orbital evolution of the binary before merger proceeds as expected, with significant precession of the spin orientation and of the orbital plane even
for the relatively short simulations considered here. Fig.~\ref{fig:prec} shows the evolution of the inclination of the black hole spin with respect
to the initial direction of the total angular momentum of the system, and the phase of the precession of the black hole spin. The first is nearly
constant, while the second shows that for the longest simulation studied here (M7-S9-I20), we observe
about three-quarters of a precession period. The orientation of the orbital plane, which we do not show here, precesses at the same frequency as the black
hole spin. These effects can be entirely understood from post-Newtonian predictions for the evolution of point particles, as already shown for
simpler equations of state~\cite{Foucart:2010eq}.

Among the simulations considered here, one falls within the range of parameters for which we expect direct plunge of the neutron star into the
black hole (M7-S7-I60), and the others are expected to disrupt (although M7-S9-I60 is close to the limit between disrupting and plunging neutron 
stars)~\cite{Foucart2012}.
We find indeed that, at the accuracy at which we can resolve the merger, there is no significant amount of matter remaining outside of the black hole
after the merger of model M7-S7-I60. At our highest resolution, the merger leaves $0.004M_\odot$ of material outside the black hole (including $0.002M_\odot$
of unbound material). At lower resolution, we found higher values of the remaining mass (a few percents of a solar mass). 
Our results are consistent with convergence of the remnant baryon mass to zero, although we cannot rule out a remnant disk and unbound ejecta at the level 
of $\sim 0.001M_\odot$. The other systems show the expected disruption of the neutron star. 
The higher mass systems, however, disrupt as the neutron star plunges into the black hole, leading to a very dynamical disruption process. 

\begin{figure}
\begin{center}
\includegraphics*[width=0.45\textwidth]{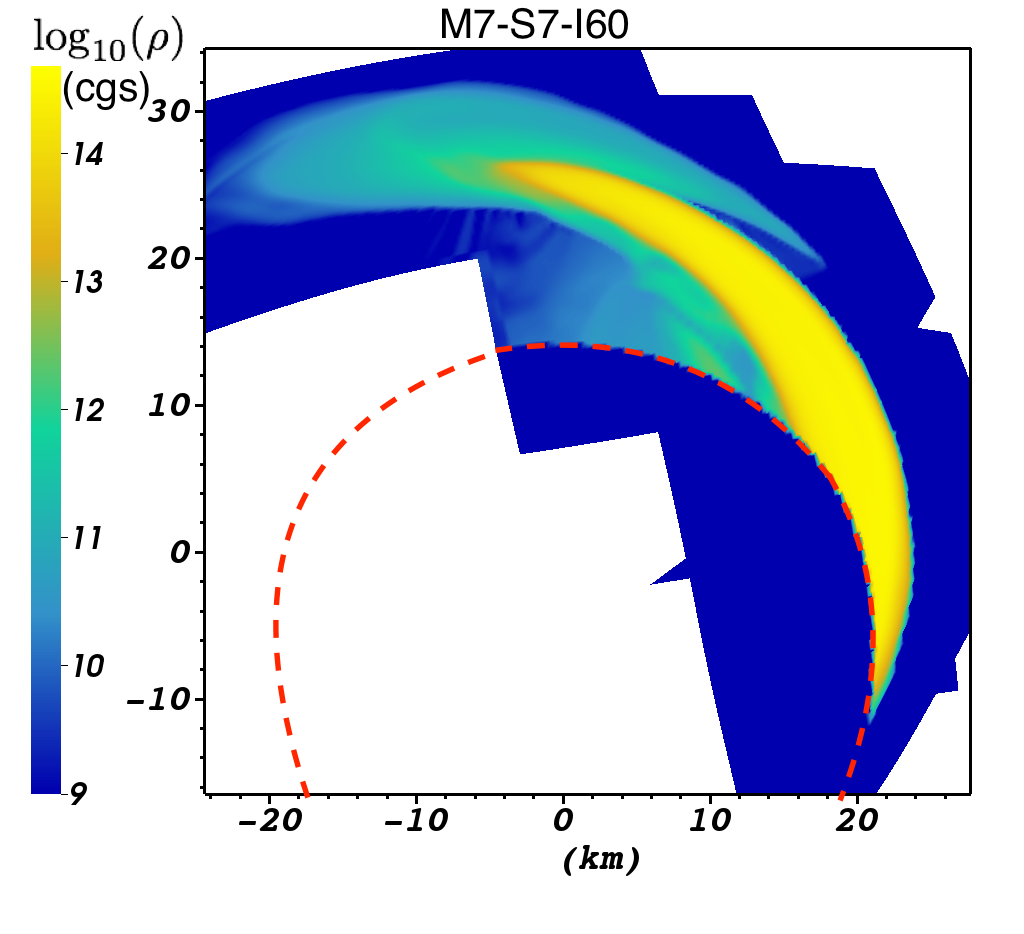}
\includegraphics*[width=0.45\textwidth]{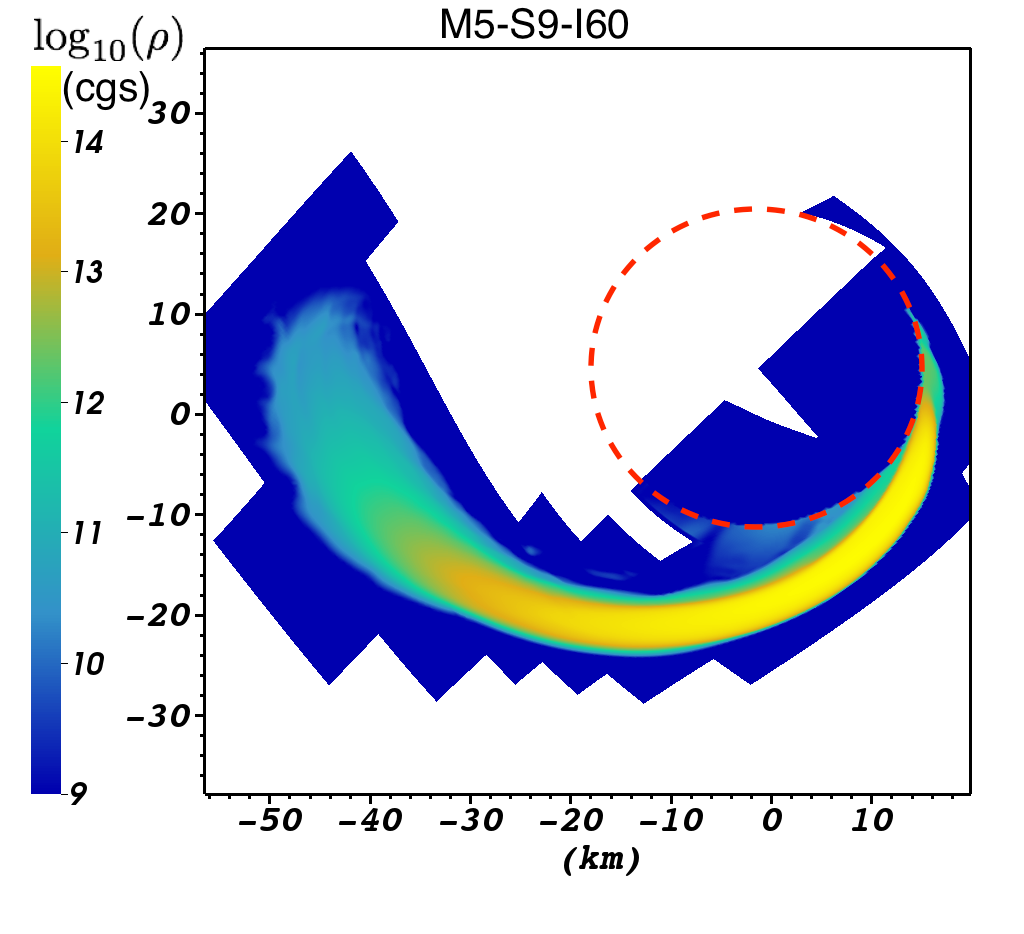}
\caption{Barion density in the orbital plane of the merging binary at the time at which half the neutron star material has been accreted by the black hole.
{\it Left}: Model M7-S7-I60. {\it Right}: Model M5-S9-I60. White regions are not covered by the finite volume grid, while the dashed red line shows the boundary
of the excised region inside the black hole. Note the different length scales used
for each plot. In model M7-S7-I60, the neutron
star plunges fully into the black hole. In model M5-S9-I60, a very narrow tidal tail forms, with a small amount of unbound material.}
\label{fig:disrupt}
\end{center}
\end{figure}

Fig.~\ref{fig:disrupt} shows a snapshot of the baryon density for the two extreme models M7-S7-I60 and M5-S9-I60. The first leads to direct plunge of the neutron
star into the black hole, while the second results in the ejection of a small amount of material ($M_{\rm ej}\approx 0.014M_\odot$, see below) and the formation
of an accretion disk. The narrow tail with sharp edges which is typical of these NSBH mergers, is clearly visible for the lower mass system M5-S9-I60. We note that,
while M5-S9-I60 is the most favorable system for disruption of the neutron star outside the ISCO, the disruption is still occurring very close to the black hole,
and partially mixes with the plunge of the neutron star into the black hole. We can also see that, during tidal disruption, the code makes full use of the adaptive 
finite volume grid discussed in~\ref{sec:grid} (white regions are not covered by the numerical grid). The formation of a narrow tidal tail is one of the most striking
difference between simulations using ideal gas or piecewise polytropic equations of state, and those using nuclear-theory, composition dependent equations of state, 
and was already noted in simulations of spin-aligned NSBH mergers~\cite{Foucart:2014nda}. For an ideal gas equation of state, the disruption of the neutron star is
generally more gradual, and the tidal tail significantly wider, as shown in Fig.3 of Foucart et al. (2014)~\cite{Foucart:2014nda}.

The main properties of the post-merger remnant and characteristics of the dynamical ejecta are summarized in Table~\ref{tab:results}, and discussed in more
detail below (except for the inaccurate model M7-S9-I60, discussed in~\ref{sec:errors}).

\begin{table}
\begin{center}
\caption{Properties of the dynamical ejecta and post merger remnant. $M_{\rm BH}^f$ and  $\chi_{\rm BH}^f$  are the mass and dimensionless spin of the black hole,
and $M_{\rm out}^f$ is the baryon mass remaining outside of the black hole. Those quantities are measured at the first minima of the accretion rate onto the black hole,
before circularization of the accretion disk. The baryon mass outside of the black hole immediately after disk formation (which is a more vaguely defined
time) is typically $10\%-20\%$ lower than $M_{\rm out}^f$. $M_{\rm ej}$ is the mass of the dynamical ejecta, and $\langle v/c\rangle_{\rm ej}$, 
$\langle i \rangle_{\rm ej}$ are the mass-weighted average of its velocity and orbital inclination with respect to the
equatorial plane of the black hole. All these properties are nearly constant, from about $1\,{\rm ms}$ after the merger. Bracketed numbers for
$M_{\rm out}^f$ and $M_{\rm ej}$ show semi-analytical predictions for the mass outside of the black hole $10\,{\rm ms}$ after merger~\cite{Foucart2012},
and the ejected mass~\cite{Kawaguchi:2016}, while bracketed numbers for $M_{\rm BH}^f$ and $\chi_{\rm BH}^f$ are semi-analytical predictions
from~\cite{Pannarale:2014}. The analytical values provided in this table for the final mass and spin of the black hole were kindly provided
to us by Francesco Pannarale.}
\label{tab:results}
{
\begin{tabular}{c|cc|cccc}
Model & $M_{\rm BH}^f\,(M_\odot)$ & $\chi_{\rm BH}^f$ & $M_{\rm out}^f\,(10^{-2} M_\odot)$ & $M_{\rm ej}\,(10^{-2} M_\odot)$ & $\langle v/c \rangle_{\rm ej}$ & $\langle i \rangle_{\rm ej}$\\
\mr
M5-S7-I60 & 6.12 [6.2] & 0.76 [0.76] & 15 [10] & 1.4 [0.6] & 0.16 &  $26^\circ$\\
M5-S9-I60 & 6.06 [6.1] & 0.86 [0.86] & 21 [19] & 1.4 [1.1] & 0.15 &  $27^\circ$\\
M7-S7-I60 & 8.22 [8.3] & 0.75 [0.75] & $\leq 0.5$ [0] & $\leq 0.25 [0]$  &  -- & --\\
M7-S9-I20 & 7.75 [8.0] & 0.90 [0.93] &  47 [31] &  4.3 [4] & 0.20 & $12^\circ$ \\
\end{tabular}
}
\end{center}
\end{table}

\subsection{Ejecta properties}
\label{sec:ejecta}

In NSBH mergers, there are effectively two sources of ejecta. The first is the {\it dynamical ejecta}, which is produced by the tidal disruption of the neutron star,
and the second is {\it disk outflows}, which are due to magnetically-driven or neutrino-driven winds in the post-merger accretion disk, and to heating of the remnant
disk due to nuclear recombination. Our merger simulations
can only constrain the mass and properties of the dynamical ejecta. We should however keep in mind that $5-20\%$ of the remaining accretion disk may be unbound
on longer timescales, thus contributing to the total amount of ejected material~\cite{Fernandez2013,Just2014}.

In our models, the mass of the dynamical ejecta vary in the range $(0.01-0.05) M_\odot$, with the high spin, high black hole mass systems producing the most
ejecta (see Table~\ref{tab:results}). This is generally less mass than for aligned-spin systems at the same mass ratio and black
hole spin~\cite{Foucart:2014nda,Kawaguchi2015}. Indeed, tidal disruption and the ejection of material are usually favored by a higher value
of the aligned component of the black hole spin. 

We note that the impact of the spin on the ejected mass is large only when the neutron star disrupts
close to the ISCO. Indeed, the main impact of a high spin is to bring the ISCO closer to the black hole, thus letting the neutron
star disrupt deeper into the gravitational potential of the black hole.
If even at low black hole spin the disruption of the neutron star occurs outside of the ISCO, the dynamics of the merger are not affected as much by the black hole spin. 
This is why the impact of
the black hole spin on the mass of the ejecta is large for $M_{\rm BH}=7M_\odot$, but small for $M_{\rm BH}=5M_\odot$. In the latter
case, the ejected material is unbound at a separation at which spin effects are relatively minor. Disk formation, which occurs later in the disruption
of the neutron star, is significantly affected by the black hole spin for all black hole masses considered here.

\begin{figure}
\begin{center}
\includegraphics*[width=0.48\textwidth]{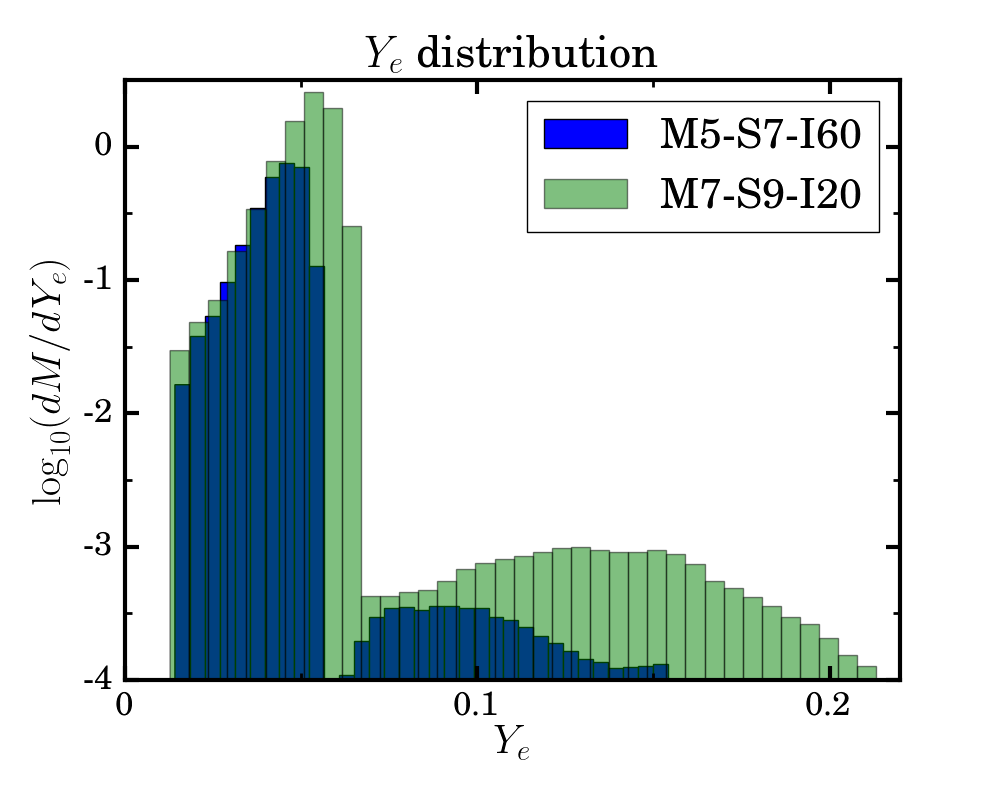}
\includegraphics*[width=0.48\textwidth]{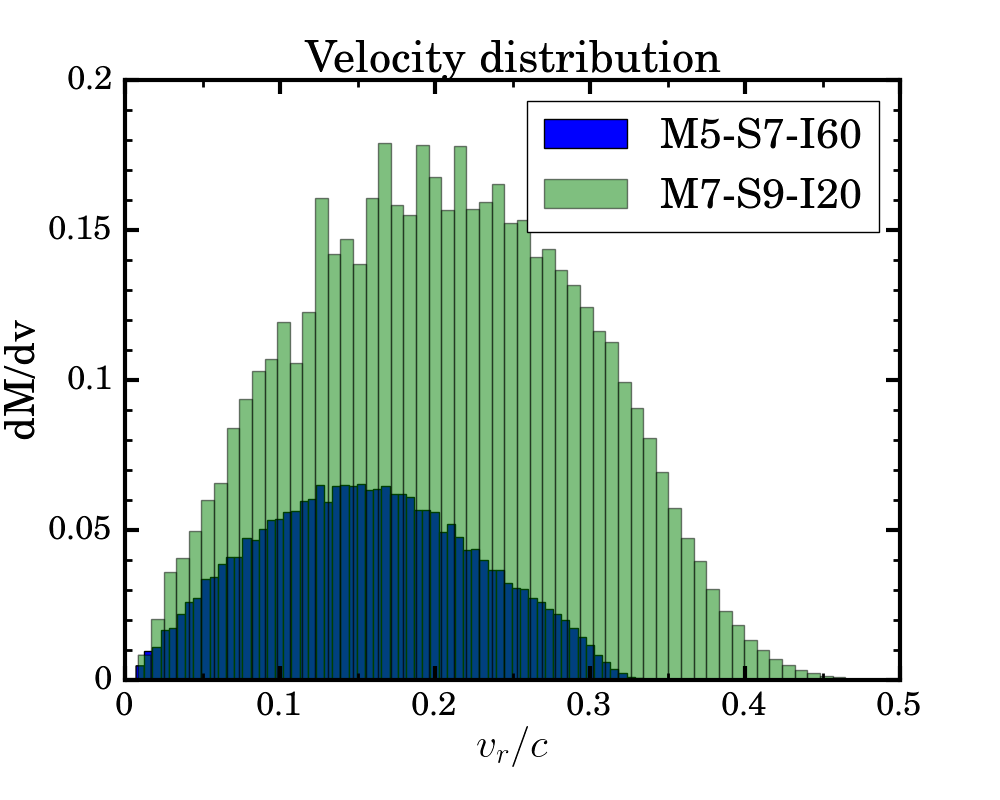}
\caption{Distribution of the electron fraction $Y_e$ ({\it Left}) and asymptotic velocity ({\it Right}) in the ejecta of models M5-S9-I60 and M7-S9-I20.
Most of the ejected material has $Y_e\leq0.06$ and all of it has $Y_e\leq 0.25$. We observe a broad velocity distribution, symmetric around
the average values $\langle v \rangle = (0.15,0.20)$.}
\label{fig:ejhist}
\end{center}
\end{figure}

As is typical for the dynamical
ejecta of NSBH mergers, the unbound material is very neutron rich. The density-weighted average electron fraction of the ejecta
is $\langle Y_e \rangle = 0.04-0.06$, depending on the simulation. For such neutron-rich ejecta, the outcome of r-process
nucleosynthesis is guaranteed to be the robust production of heavy elements, including the 2nd and 3rd peak of the r-process,
as well as high-opacity lanthanides, with little to no production of elements around the observed 1st peak of the 
r-process~\cite{Wanajo2014,Lippuner2015,Roberts:2016}. 
A more detailed description of the composition of the ejecta for two of our models is presented in Fig.~\ref{fig:ejhist} (left panel). 
We see that there is no significant amount
of material with $Y_e>0.07$ (note the logarithmic scale used for this plot), and absolutely no observed ejecta at $Y_e>0.25$,
where lighter r-process elements could be produced. Overall, we can thus conclude that (i) r-process nucleosynthesis in this dynamical ejecta will only produce
the heavier r-process elements~\cite{Lippuner2015}, and (ii) the opacity of the ejecta will be high, causing the electromagnetic transient
powered by the r-process to peak in the infrared~\cite{2013ApJ...775...18B}.

The asymptotic velocity of the ejecta is another important parameter if we want to predict the properties of that infrared transient.
Larger velocities imply faster expansion of the ejecta, and a larger amplitude, shorter lived transient~\cite{2013ApJ...775...18B,Barnes:2016}. 
The right panel of 
Fig.~\ref{fig:ejhist} shows the distribution
of velocities in the ejecta of two of our models. The density-weighted
average velocity of the ejecta is $\langle v\rangle \sim 0.15c$ for $M_{\rm BH}=5M_\odot$ and $\langle v \rangle \sim 0.20c$ for $M_{\rm BH}=7M_\odot$. 
We see that the ejecta has a broad velocity distribution, covering the range $0\leq v \leq 2\langle v\rangle$. The measured kinetic energy
of the ejecta ranges from $3\times 10^{50}\,{\rm ergs}$ for model M5-S9-I60 to $1.7\times 10^{51}\,{\rm ergs}$ for M7-S9-I20. The kinetic energy
of the ejecta plays an important role as the available amount of energy to power radio transients as the ejecta interacts with the interstellar
medium~\cite{Nakar:2011cw}. We note that the typical velocity of the ejecta measured here is comparable to what was obtained in spin-aligned simulations
at the same mass ratio using the hot, composition dependent nuclear-theory based equation of state LS220~\cite{Foucart:2014nda}, but is lower than in simulations
using a fixed fluid composition~\cite{Foucart:2013a,Kawaguchi2015}. We discuss the origin of this difference in Sec.~\ref{sec:discussion}.

\begin{figure}
\begin{center}
\includegraphics*[width=1.\textwidth]{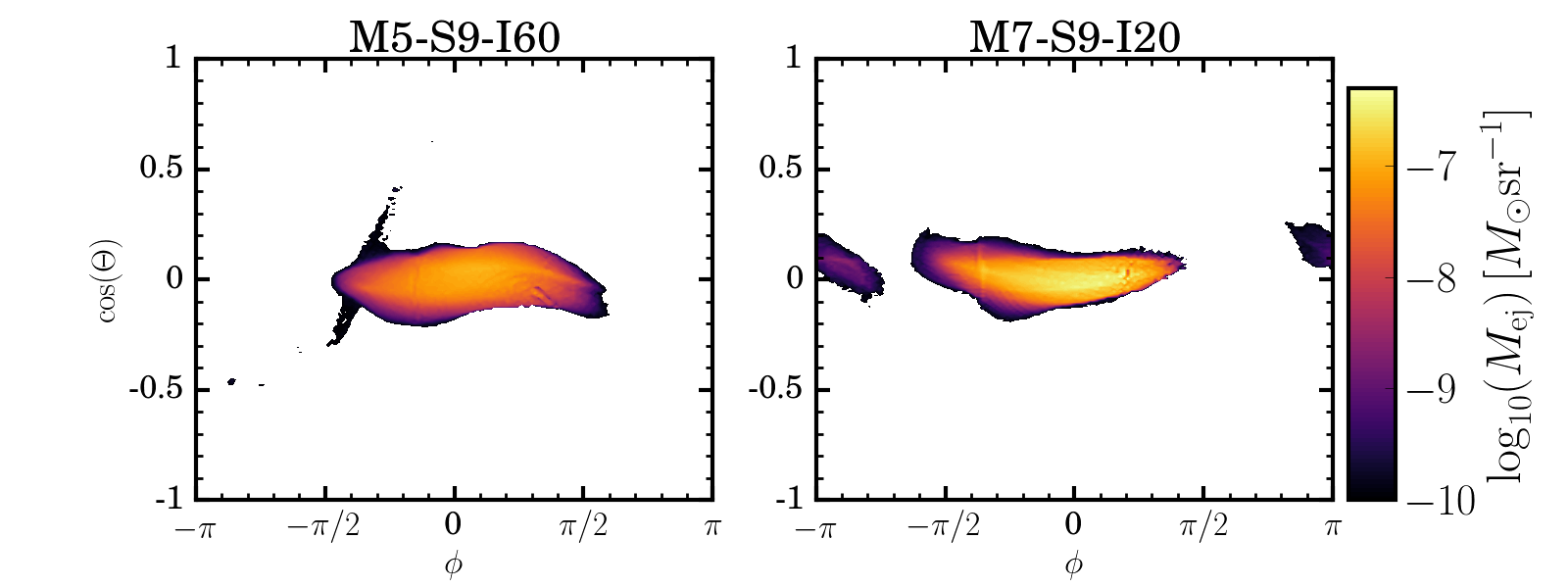}
\caption{Angular distribution of the ejected material for simulations M5-S9-I60 ({\it Left}), and M7-S9-I20 ({\it Right}).
We use spherical polar coordinates centered on the black hole, with the polar axis chosen so that the equatorial plane
is the best fit plane for the orbital plane of the unbound material. The ejected material fills a narrow region of opening angle $\sim 20^\circ$
($\sim 180^\circ$) in the vertical (azimuthal) direction.}
\label{fig:ejangle}
\end{center}
\end{figure}

The geometry of the ejecta can also have important effects on the properties of kilonovae, mainly by introducing a dependence of the kilonova
signal in the orientation of the binary with respect to the observer. In NSBH mergers, this is particularly important because the dynamical
ejecta is very opaque and neutron rich, while material ejected later in disk outflows could have a high enough electron fraction to prevent
the production of heavy r-process elements, thus resulting in a less opaque ejecta, and a bluer, brighter, and shorter electromagnetic 
signal~\cite{Fernandez2013,Just2014,Perego2014,Fernandez:2014,Fernandez:2014b}.
More precisely, this is more likely to happen for disk outflows in the polar regions of the post-merger remnant, 
and requires that the disk outflows are not masked by high-opacity dynamical ejecta. 

In Fig.~\ref{fig:ejangle}, we show the angular distribution of the dynamical
ejecta, in a spherical polar coordinate system $(r,\theta,\phi$) chosen so that the average orbital plane of the ejecta corresponds to $\theta = 0$. The inclination
$\langle i \rangle_{\rm ej}$ of that coordinate system with respect to the black hole spin is listed in Table~\ref{tab:results}. We see that the overall geometry
of the dynamical ejecta is consistent with the results of previous NSBH merger simulations~\cite{kyutoku:2015}. The ejecta forms a crescent with an azimuthal opening 
angle of $\sim 180^\circ$, and a vertical opening angle of $\sim 20^\circ$. In all of our models, the misalignment between the orbital plane of the ejecta
and the equatorial plane of the spinning black hole remains small ($\leq 30^\circ$), and the dynamical ejecta should thus be unable to obscure emission
from polar disk outflows. While differential precession causes a slight warp of the orbital plane of the ejecta, this remains a very small effect. This is due to the fact that
the ejecta rapidly leaves the strong field region in which orbital precession is significant.

\begin{figure}
\begin{center}
\includegraphics*[width=1.\textwidth]{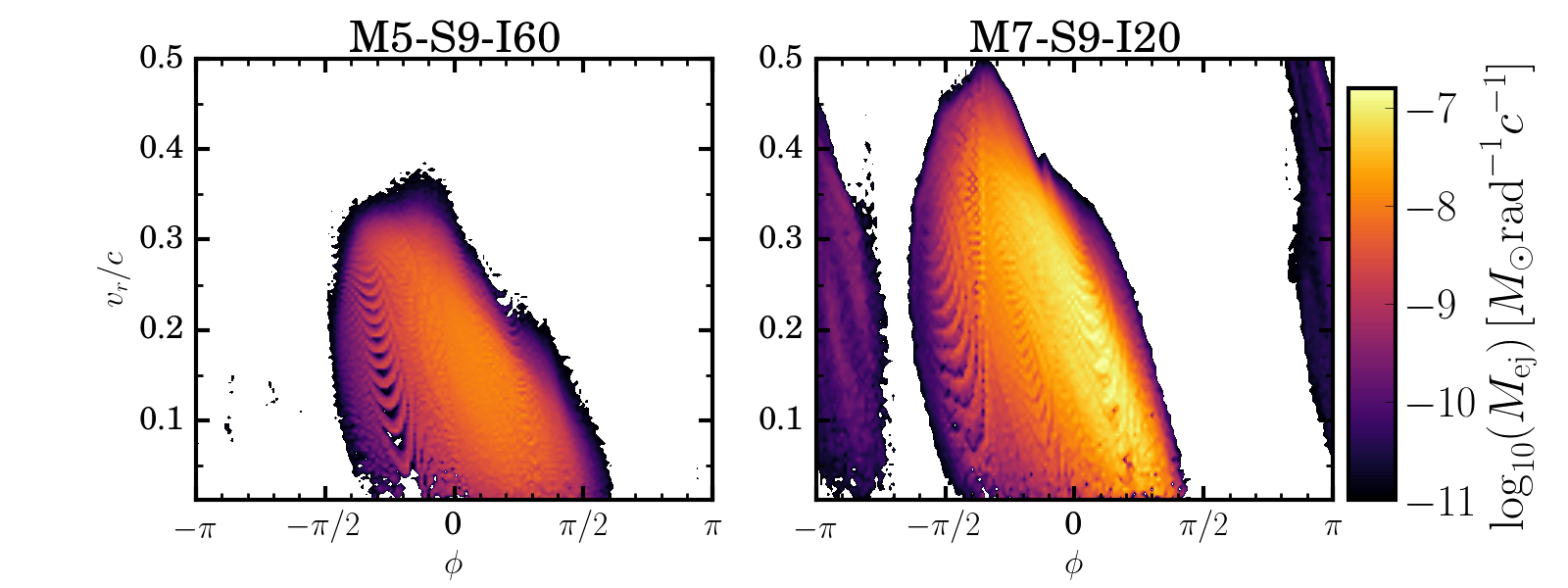}
\caption{Distribution of the asymptotic radial velocity of the ejected material for simulations M5-S9-I60 ({\it Left}), and M7-S9-I20 ({\it Right}),
as a function of the azimuthal angle $\phi$ used in Fig.~\ref{fig:ejangle}. There is a clear linear trend in the evolution of the average velocity
with azimuthal angle, which may create significant variations in the timescale of kilonovae as a function of the orientation of the binary with
respect to the observer. Material at low $\phi$ is ejected earlier and moving faster. We can approximate the bounds on the velocity $v(\phi)$
as $v_{\rm min}=\max\left[-\frac{4\phi}{\pi}\langle v\rangle,0\right]$, $v_{\rm max}=\min\left[2\left(1-\frac{2\phi}{\pi}\right)\langle v\rangle,2\langle v\rangle\right]$.
The ripples observed in velocity space are most likely of numerical origin.}
\label{fig:ejvel}
\end{center}
\end{figure}

We also note that there exist a potentially important correlation between the velocity of the ejecta and the azimuthal angle $\phi$, shown
on Fig.~\ref{fig:ejvel}. Not surprisingly, ejecta farthest from the bound material in the azimuthal direction is also given more kinetic energy and
has a larger asymptotic velocity. The breadth of the ejected material in the radial direction will be smaller with this effect taken into account
that if one assumes a single broad velocity distribution (as in e.g.~\cite{Kawaguchi:2016}). The impact of this effect on kilonova lightcurves will have to
be tested using 3D radiation transport. Fig.~\ref{fig:ejvel} indicates that a rough estimate of the range of velocity $(v_{\rm min},v_{\rm max})$
observed at an azimuthal angle $\phi$ is
\beqn
v_{\rm min} &=& \max\left[-\frac{4\phi}{\pi}\langle v\rangle,0\right],\\
v_{\rm max}&=& \min\left[2\left(1-\frac{2\phi}{\pi}\right)\langle v\rangle,2\langle v\rangle\right],
\eeqn
with $\phi$ in the range $[-\pi/2,\pi/2]$ covering the regions in which a significant amount of ejecta is present. If the velocity distribution significantly
affects our ability to extract useful information about the merging objects from the observation of kilonovae, a more accurate model will certainly
have to be designed based on a larger number of simulations (including some ejecting a small amount of material in 
a larger arc $\delta \phi \sim 2\pi$, as observed in~\cite{kyutoku:2015}).

\subsection{Post-merger remnant}
\label{sec:remnant}

\begin{figure}
\begin{center}
\includegraphics*[width=0.52\textwidth]{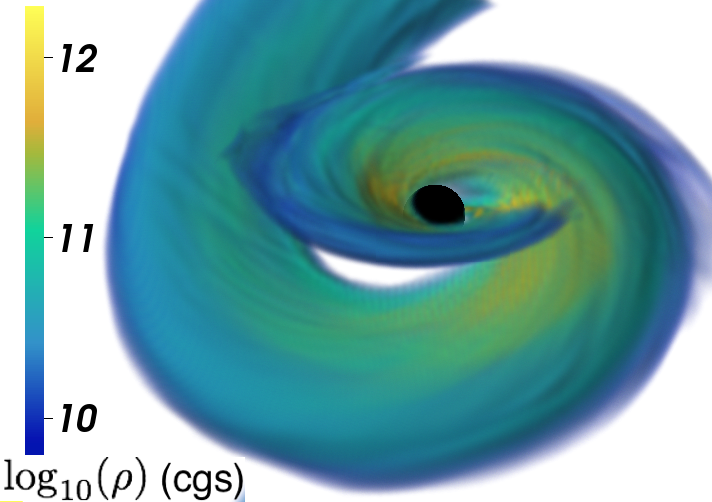}
\includegraphics*[width=0.44\textwidth]{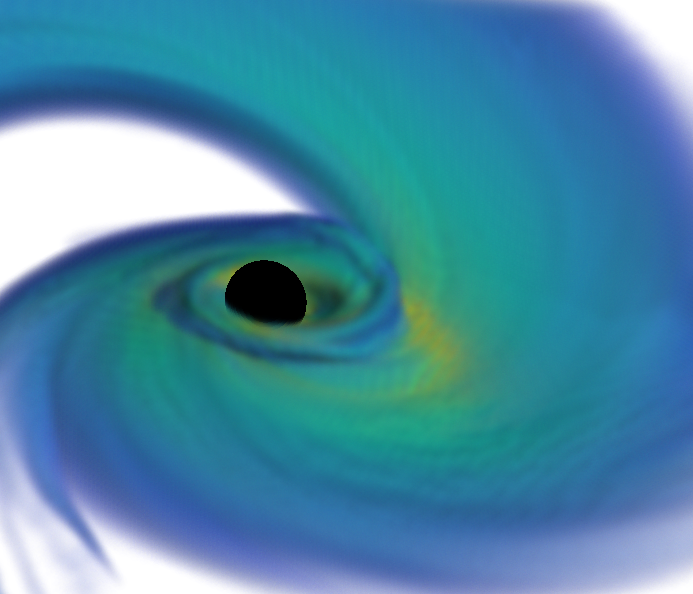}
\caption{Baryon density at the beginning of disk circularization ($2.5\,{\rm ms}$ after merger) ({\it Left}), and $3\,{\rm ms}$ later ({\it Right}), 
for model M5-S7-I60. The formation of a wide, hot disk is slow compared to the dynamical timescale of the inner disk: $5.5\,{\rm ms}$ after merger, the peak of the
surface density of the fluid is still at $r_{\rm peak}\sim 30\,{\rm km}$, while the horizon of the black hole has a radius $r_{\rm BH}\sim 27\,{\rm km}$.
The system remains closer to an accreting tidal tail than a circularized disk.
}
\label{fig:diskform}
\end{center}
\end{figure}

Closer to the black hole, the disrupted neutron star forms a precessing tidal tail, which eventually leads to the formation of a disk over a timescale
of $(5-10)\,{\rm ms}$. Rapid accretion onto to black hole stops $\sim 2-3\,{\rm ms}$ after merger (we define the merger time as the time at which half the neutron star mass
has been accreted onto the black hole). At that time, the accretion rate and average temperature reach clear minima. 
As this is a convenient feature of all simulations which
immediately follows the merger, we choose that time to measure many of the quantities listed in Table~\ref{tab:results} (mass and spin of the black hole, baryon mass
remaining outside of the black hole). 
The decrease in accretion rate and temperature stops as the tidal tail self-intersects and shocks (see left panel of Fig.~\ref{fig:diskform}, for model M5-S7-I60). 
This self-interaction of the tail material begins a slow circularization of the tidal tail into an accretion disk. The right panel of Fig.~\ref{fig:diskform} shows model
M5-S7-I60 $5.5\,{\rm ms}$ after merger. At that time, the disk remains very far from axisymmetry, and the surface density still peaks right on the horizon of
the black hole. 

To observe the formation of an accretion disk and its early evolution, we continue the evolution of model M5-S7-I60 up to $17\,{\rm ms}$ after merger.
The true accretion disk really forms $\sim 10\,{\rm ms}$ after merger, with a peak of the mass distribution at a radius $r_{\rm disk}\sim (50-60)\,{\rm km}$,
a temperature $T\sim 3-6\,{\rm MeV}$, and an electron fraction $Y_e\sim 0.1-0.2$. The disk properties are thus very similar to those of non-precessing systems
evolved with a different hot, nuclear-theory based equation of state~\cite{Foucart:2014nda}. 
During disk formation, the accretion rate on the black hole remains very high, with
$\dot M_{\rm BH}\sim 5M_\odot {\rm s}^{-1}$ on average. The inner regions of the disk rapidly align with the equatorial plane of the black hole, as already observed
for simpler equations of state~\cite{Kawaguchi:2015}. 
While the tidal tail is initially at a $20^\circ-25^\circ$ inclination with respect to the equatorial plane of the back hole, that
misalignment drops to $\sim 10^\circ$ only $10\,{\rm ms}$ after merger. 
This change is not due to accretion (the black hole only accretes $\sim 0.03M_\odot$ during
this period), and thus has to be due to a combination of hydrodynamical shocks 
and/or non-linear damping of the inclination due to the interaction between neighboring regions of the differentially precessing disk (see e.g.~\cite{Sorathia:2013}).

The more violent disk formation phase is followed by a slower secular evolution of the remnant. For the entire duration of the simulation, the accretion rate onto
the black hole remains high, at $\dot M_{\rm BH}\sim 5M_\odot {\rm s}^{-1}$. A large fraction of the matter outside of the black hole remains in the bound tidal tail.
 The accretion disk, which we arbitrarily define as the matter within $\sim 150\,{\rm km}$ of the black hole (a convenient choice because the cumulative mass 
 distribution is nearly flat for $100\,{\rm km}<r<300\,{\rm km}$), has a mass decreasing from $M_{\rm disk}=0.06M_\odot$ to $M_{\rm disk}=0.045M_\odot$ 
 between $10\,{\rm ms}$ and $17\,{\rm ms}$ after the merger. This leaves a bound tidal tail of mass $M_{\rm bound}=0.027M_\odot$, as well as an ejected mass
 of $M_{\rm ej}=0.014M_\odot$ at the end of the simulation. The observed accretion rate can obviously not be sustained for more than another $\sim 10\,{\rm ms}$.
 The fallback rate is thus expected to quickly become an important driver of the evolution of the disk. 
 
 We note that with such disk masses, the mass ejected
 through disk outflows is expected to be only a factor of a few smaller than the mass of the dynamical ejecta, with large uncertainties in both the mass and properties
 of the outflows. This should be the case for all of the configurations studied here, and indicates that those disk outflows could contribute to the
 production of a radioactively powered transient.
 
 The temperature of the disk remains constant over the duration of the simulation, despite neutrino luminosities of $L_\nu \sim (1.7-3.8) \times 10^{53}{\rm erg/s}$,
 and the fact that we do not include magnetic fields and thus do not benefit from heating due to magnetic turbulence. In our simulations, the heating is entirely
 driven by hydrodynamical shocks during the circularization of the material falling back onto the disk. Similarly, after a phase of rapid protonization of the fluid during
 disk formation, the electron fraction of the disk remains nearly constant. While the disk continuously emits more electron antineutrinos than electron neutrinos,
 that effect is mostly compensated by the accretion of high-$Y_e$ material by the black hole, and the fallback of low-$Y_e$ material from the tidal tail.
 The relatively constant properties of the post-merger remnant indicate that snapshots of our simulation anywhere in the range $(10-17)\,{\rm ms}$ after
 merger probably provide reasonable initial conditions for long-term
 evolutions of these disks, during which magnetically-driven turbulence and neutrino transport will play a more important role.
 
 \begin{figure}
\begin{center}
\includegraphics*[width=0.4\textwidth]{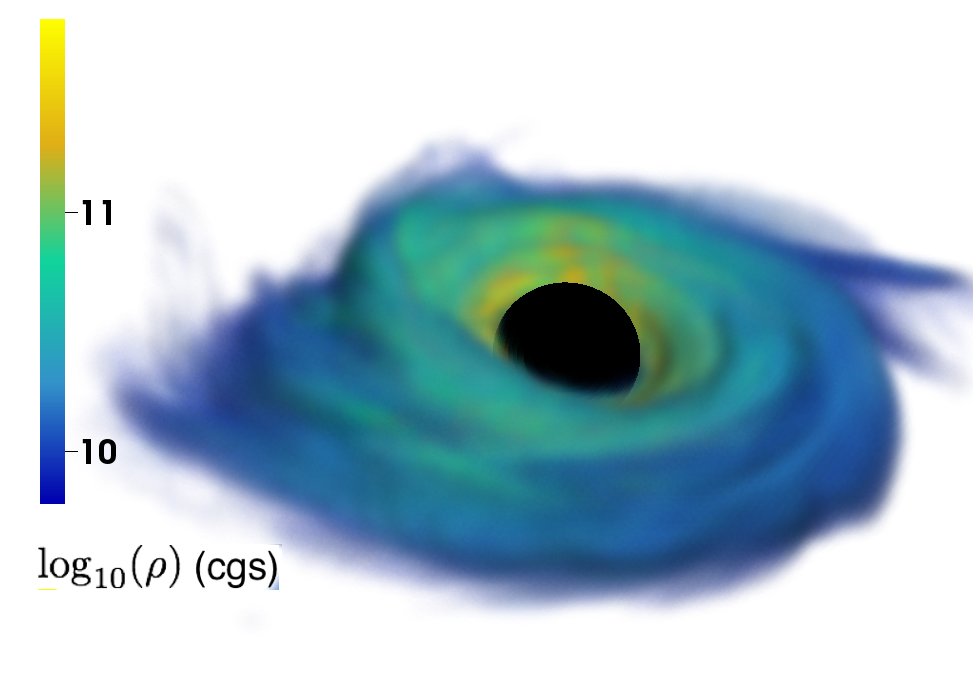}
\includegraphics*[width=0.56\textwidth]{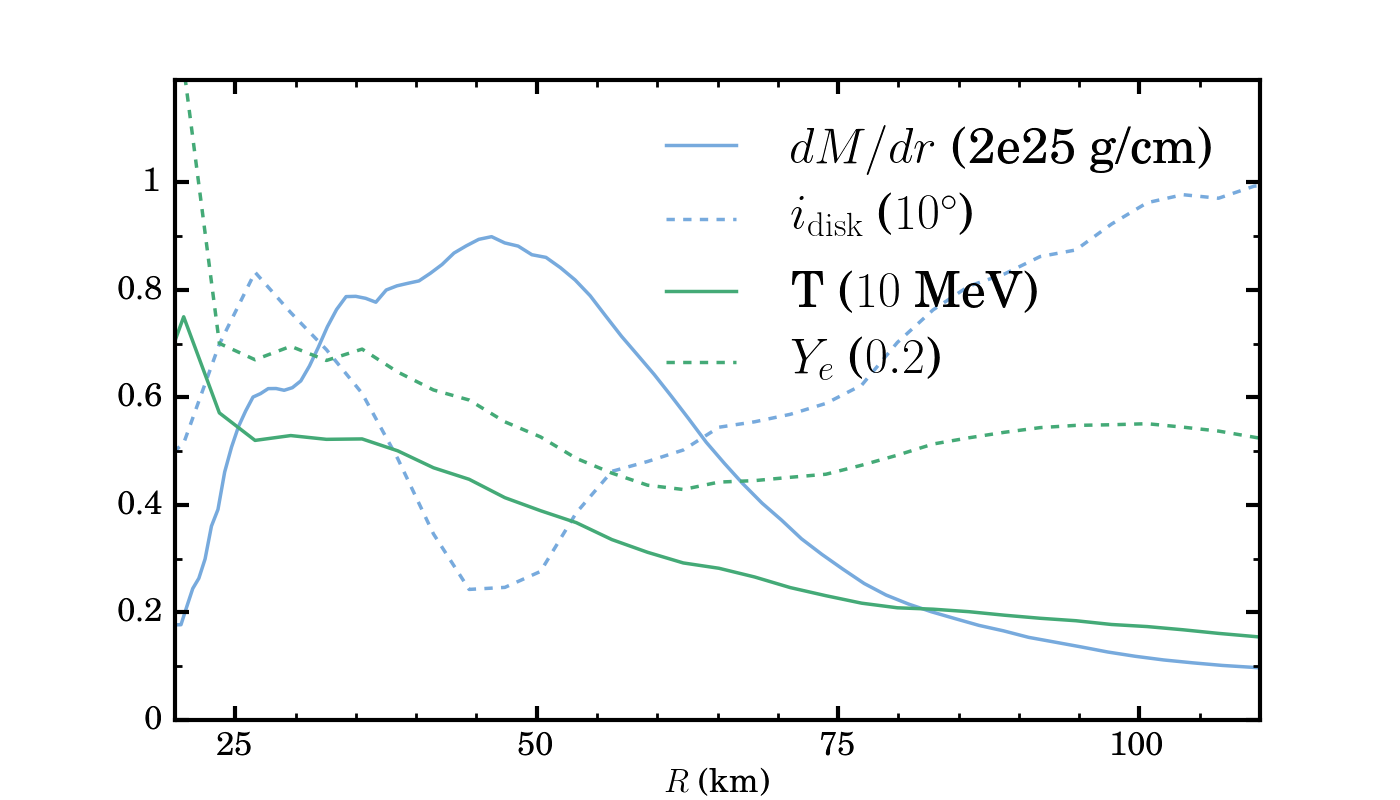}
\caption{{\it Left}: Baryon density $17\,{\rm ms}$ after merger for model M5-S7-I60. {\it Right}: One-dimensional profile of the linear distribution of the baryon mass
($dM/dr$), temperature ($T$), electron fraction ($Y_e$), and inclination of the disk angular momentum with respect to the black hole spin ($i_{\rm BH}$), 
at the same time. Number in parenthesis indicate the scale used for each quantity.}
\label{fig:disk}
\end{center}
\end{figure}
 
A more detailed view of the final disk is provided in Fig.~\ref{fig:disk}. We see that the disk is very narrow, with most of the matter within $25\,{\rm km}<r<75\,{\rm km}$.
There also remains a mild variance of $\sim 4^\circ$ in the orientation of the disk. Otherwise, the disk has settled into a configuration similar to what is observed in
other NSBH mergers, with a narrow annulus of nearly axisymmetric material, an inverted temperature gradient (with the inner edge hotter than the outer edge),
and a disk height $H\sim 0.2 r$.

Finally, we monitor the evolution of the disk inclination over the entire evolution. We note that the alignment of the angular momentum of the disk with 
 the spin of the black hole continues for the entire duration of the evolution, despite significant accretion of misaligned fallback material. The misalignment angle
 between the angular momentum of the disk (within a narrow annulus around its density peak) and the spin of the black hole continuously drops from $10^\circ$ about
 $10\,{\rm ms}$ after merger to about $3^\circ$ at the end of the simulation. There is no noticeable global precession of the disk during that period. The alignment
 of the disk occurs much faster than in our previous simulations of precessing systems~\cite{Foucart:2010eq}, presumably due to the larger black hole mass,
larger black hole spin, and smaller disk mass in the simulations presented here.

\section{Comparison with existing fitting formulae}
\label{sec:discussion}

In recent years, analytical formulae fitted to numerical simulations have been proposed to predict the baryon mass remaining outside
of the black hole after NSBH mergers~\cite{Pannarale:2010vs,Foucart2012}, the mass and velocity of the ejected 
material~\cite{Kawaguchi:2016}, and the final mass and spin of the black hole~\cite{Pannarale:2014}. 
In Table~\ref{tab:results},
we provide as references the predictions of numerical fits to the remnant baryon mass from~\cite{Foucart2012} (extended to misaligned black hole spins
as suggested in~\cite{2013PhRvD..87h4053S}), the ejected mass 
from~\cite{Kawaguchi:2016}, and the final black hole properties from~\cite{Pannarale:2014}. We see good 
qualitative agreements between the analytical predictions and our numerical results. For the low mass systems, the larger remnant masses observed in the simulations
 are simply due
to difference in the time at which the remnant mass is measured (e.g. model M5-S7-I60 has a remnant mass outside of the black hole of $0.12M_\odot$ if
we measure it $10\,{\rm ms}$ after merger, as for the fitting formula~\cite{Foucart2012}). For higher mass systems with high spins, the fitting formula somewhat
underestimates the remnant mass. This is a common issue when the remnant mass is large (see e.g.~\cite{Lovelace:2013vma}). 
Yet the results remain qualitatively correct, and in 
particular both the simulations and the fit predict that model M7-S7-I60 does not lead to the disruption of the neutron star, while all other models do.
The fit to the final mass and spin of the black hole~\cite{Pannarale:2014} generally performs as well or better than expected. As that fit
mostly relies on the knowledge of the baryon mass remaining
outside of the black hole~\cite{Foucart2012} and on conservation of energy and angular momentum, the absolute error in the final mass is comparable to the error in the
remnant baryon mass. The spins are very accurately predicted for 3 out of the 4 models, while we observe a slightly larger than expected error (0.03) for the final spin
of model M7-S9-I20.

The ejected mass is also accurately predicted by the existing fitting formula, at least within the scatter of $\sim 0.01M_\odot$ also observed in
 the simulations used to calibrate the model~\cite{Kawaguchi:2016}. 
The average velocity shows more significant differences. According to the fitting formula, we
would expect an average (root-mean-square) velocity of $\langle v_{\rm ej,rms}\rangle=(0.245,0.267)c$ for black hole masses of $M_{\rm BH}=(5,7)M_\odot$, with 
$\langle v_{\rm ej,rms}\rangle = \sqrt{2E_{\rm kin}/M_{\rm ej}}$, and $E_{\rm kin}$ the kinetic energy of the ejecta. With the same definition, 
we have $\langle v_{\rm ej,rms}\rangle=(0.175,0.22)c$ in our simulations. 
The difference between $\langle v_{\rm ej,rms}\rangle$ and the values of $\langle v\rangle$ listed in Table~\ref{tab:results} is due to the 
change of definition of the average velocity. $\langle v_{\rm ej,rms}\rangle$ is the more relevant definition when considering the kinetic energy in the ejecta, which
determines e.g. the amount of energy of the radio transients powered by interactions of the ejecta with the interstellar medium~\cite{Nakar:2011cw}, 
while $\langle v_{\rm ej} \rangle$ determines the expansion velocity of the ejecta, which has important effects on the duration and luminosity of 
kilonovae~\cite{2013ApJ...775...18B,Barnes:2016}.
Even accounting for this difference in the definition of the average energy, however, we are still left with velocities $(20-40)\%$ lower than predicted by the fitting 
formula~\cite{Kawaguchi:2016}.

There is however an important difference between the equations of state considered in the simulations used to calibrate the fitting formula~\cite{Kawaguchi:2015}, 
and the equation of state used in our simulations.
The first are composition independent, and fitted to the equation of cold matter in nuclear statistical equilibrium (NSE) and beta-equilibrium. The second is composition
dependent and only assumes NSE. As the ejecta in NS-BH mergers is cold, the ejected material in our simulations expands {\it at (nearly) constant
composition}. On the other hand, for the piecewise polytropic equations of state used in~\cite{Kawaguchi:2015}, the ejected material expands {\it in beta-equilibrium}.
Looking at the internal energy of cold, low density matter in the DD2 equation of state, we find that the difference in internal energy 
between matter at $Y_e\sim 0.05$ (as in our ejecta) 
and matter at $Y_e\sim 0.5$ (the effective composition of the piecewise polytropic equations of state at low-density) is $\Delta \epsilon = 7.6\,{\rm MeV/nuc}$.
This is simply due to the fact that cold matter at $Y_e\sim 0.05$ is assumed to be formed of mostly neutrons, with $\sim 10\%$ of low-mass seed nuclei, while
matter at $Y_e\sim 0.5$ is assumed to be mostly formed of $^{56}{\rm Ni}$. Neither of these assumptions is correct: the ejecta in fact drops out of NSE 
at temperatures $T\sim 0.5\,{\rm MeV}$, and its composition is then determined by the result of r-process nucleosynthesis.

These different assumptions can account for a large part of the disagreement in the predicted velocity of the ejecta. If all of the energy which
would be released by transforming our neutron rich ejecta into $^{56}{\rm Ni}$ was converted into kinetic energy, we would 
get $\langle v_{\rm ej,rms}\rangle=(0.22,0.24)c$, which brings us within the $\sim 10\%$ relative error expected for the fitting formula.

The correct answer lies somewhere in between the two results. Nucleosynthesis in the ejecta is expected to release $\sim 6\,{\rm MeV/nuc}$.
How much of that energy is used to accelerate the ejecta depends on the efficiency with which the products of nuclear reactions during r-process nucleosynthesis
are thermalized. Thermalization in the ejecta has been studied in some detail late in the evolution of the system 
(on the timescale for kilonova emission)~\cite{Barnes:2016}, 
but not in the first second when most of the r-process heating is occurring. We will here follow~\cite{2010MNRAS.402.2771M} and make the rough
estimate that about half of the energy released is lost to neutrinos. This is a good approximation if most of the energy is released in beta-decays 
producing relativistic electrons and neutrinos. With this assumption, we can estimate that the ejecta will have a kinetic energy $\sim 4.6\,{\rm MeV/nuc}$
lower than predicted using piecewise polytropic equations of state. We can thus derive a new prediction for the average velocity of the ejecta, based on
the existing fit~\cite{Kawaguchi:2016}, by removing $\sim 4.6\,{\rm MeV/nuc}$ from the estimated kinetic energy:
\beqn
\langle v_{\rm ej,rms}\rangle &=& 0.0166*Q + 0.1657,\\
\langle v_{\rm ej}\rangle &=& 0.0149*Q + 0.1493.
\eeqn
For the second formula, we assumed a ratio of $\sim 1.11$ between the root-mean-square and linear average of the velocity, as observed in our simulations.
Once differences in the composition of the fluid at low density are taken into account, this estimate is consistent with our numerical simulations
and with pre-existing results. We note that while differences in the average velocity of $\delta \langle v \rangle \sim (0.02-0.05)c$ may seem fairly trivial, the properties
of kilonovae are very strongly affected by changes in the ejecta velocity~\cite{2013ApJ...775...18B,Barnes:2016}. 
Any attempt to extract information about the ejected mass
from an observed kilonova could be significantly biased if velocities are inaccurately estimated. They may also be affected by the spatial
distributions of velocities, which has been idealized in the only existing ejecta model~\cite{Kawaguchi:2016}. 

\section{Conclusions}
\label{sec:conclusion}

We perform the first general relativistic simulations of precessing neutron star-black hole mergers in which the neutron star matter is modeled using a hot,
temperature-dependent, nuclear-theory based equation of state. We cover a range of black hole masses expected from the observation
of galactic X-ray binaries, and black hole spins chosen to allow for tidal disruption of the neutron star in most simulations. We focus
on the dynamical ejection of material by the merger, and the formation of an accretion disk, but do not follow the long-term evolution of 
the remnant disk. Longer evolutions of the disk would require the inclusion of magnetic effects, and a better treatment of the neutrinos
than the simple leakage scheme used in this work.

We find that the disruption (or non-disruption) of the neutron star, the amount of matter ejected by the merger, and the properties of the
post-merger remnant (disk mass, black hole mass and spin) are consistent with existing fitting formulae, even those which are
extrapolated from the results of non-precessing merger simulations~\cite{Foucart2012,Pannarale:2014}. The ejected
mass is also consistent with predictions derived from precessing merger simulations using a simpler equation of 
state~\cite{Kawaguchi:2015,Kawaguchi:2016}. The velocity of the ejecta, on the other hand, differ significantly from those predictions.
We show that this disagreement can be traced to the assumptions made by the different equations of state at low density and temperature, 
which affect the evolution of the neutron-rich, cold dynamical ejecta. We argue that none of the equations of state used so far properly
capture the evolution of that ejecta, but that simple corrections to the velocities extracted from numerical simulations can easily be applied
to improve our estimates of the asymptotic ejecta velocity, and reconcile the results of simulations using different equations of state.

The geometry of the disrupting neutron star also appears to differ between composition-dependent equations of state and 
piecewise-polytropic~\cite{Kawaguchi:2015} or polytropic~\cite{Foucart:2010eq,Foucart:2013a} equations of state. The tidally disrupted star forms a narrower
tidal tail, thus making the merger significantly harder to resolve numerically (see also~\cite{Foucart:2014nda} for the same effect in non-precessing systems).
To partially offset this problem, we developed an improved method to adaptively construct our numerical grid, and only cover with the grid regions in which
matter is currently present. This new method improves our ability to resolve these mergers, even though numerical errors remain a significant issue in one
simulation of a marginally disrupting neutron star around a rapidly spinning black hole (model M7-S9-I60).

We also show that all of the dynamical ejecta produced in these mergers is very neutron-rich, and will undergo strong r-process nucleosynthesis.
The velocity distribution of the ejecta appears very similar in all of our simulations, with a broad distribution covering the range $0<v<2\langle v\rangle$,
and $\langle v\rangle$ the average velocity of the ejecta. We also provide an approximate model for the angular variation in the velocity distribution
of the ejecta, which could be helpful in constructing improved lightcurve models for kilonovae. From the post-merger disk mass and ejected mass,
we can infer that disk outflows ejecting material from the post-merger remnant over longer timescales~\cite{Fernandez2013,Just2014} are likely to 
be subdominant when compared to the dynamical ejecta, but not negligible (maybe $\sim 2-3$ times less massive, with large uncertainties on the mass
and properties of the disk outflows). As the dynamical ejecta does not obscure the polar regions of the remnant, if less neutron-rich polar disk outflows 
are generated they could power a significant optical component to the kilonova signal.

Finally, our results indicate that the remnant accretion disk rapidly aligns with the equatorial plane of the spinning black hole remnant,
as in simulations using piecewise polytropic equations of state~\cite{Kawaguchi:2015}. For $M_{\rm BH}=5M_\odot$, the bound material forms a 
narrow, hot, and compact accretion torus, with most of the mass at radii $25\,{\rm km}<r<75\,{\rm km}$. Continuous fallback onto the disk causes shocks in the post-merger
remnant which heat the disk and drive strong accretion onto the black hole ($\dot{M}\sim5M_\odot/s$), and sustained neutrino emission ($L_\nu > 10^{53}{\rm erg/s}$).
This provides ideal initial conditions for the production of hotter disk outflows and/or short gamma-ray bursts. The long-term evolution of this remnant cannot however
be reliably performed with the limited physics included in these simulations.

\ack
The authors thank Jennifer Barnes, Rodrigo Fernandez, Brian Metzger, Eliot Quataert, Sasha Tchekhovskoy, and the members of the SxS collaboration
for helpful discussions over the course of this project. We also thank Francesco Pannarale for providing information about the predicted properties of the final 
black holes, listed in Table~\ref{tab:results}. Support for this work was provided by NASA through Einstein Postdoctoral Fellowship grant numbered PF4-150122 (F.F.)
awarded by the Chandra X-ray Center, which is operated by the Smithsonian Astrophysical Observatory for NASA under contract NAS8-03060.
D.D. gratefully acknowledges support from the UC Berkeley-Rose Hills Foundation Summer Undergraduate Research Fellowship.
D.K. is supported in part by a Department of Energy Office of Nuclear
Physics Early Career Award, and by the Director, Office of Energy
Research, Office of High Energy and Nuclear Physics, Divisions of
Nuclear Physics, of the U.S. Department of Energy under Contract No.
DE-AC02-05CH11231. 
H.P. gratefully acknowledges support from the NSERC Canada. 
M.D. acknowledges support through NSF Grant PHY-1402916. 
L.K. acknowledges support from NSF grants PHY-1306125 and AST-1333129 at Cornell, 
while the authors at Caltech acknowledge support from NSF Grants PHY-1404569, AST-1333520, NSF-1440083, and NSF CAREER Award PHY-1151197. 
Authors at both Cornell and Caltech also thank the Sherman Fairchild Foundation for their support.
Computations were performed on the supercomputer Briar\'ee from the Universit\'e de Montr\'eal, 
managed by Calcul Qu\'ebec and Compute Canada. The operation of these supercomputers is funded
by the Canada Foundation for Innovation (CFI), NanoQu\'ebec, RMGA and the Fonds de recherche du Qu\'ebec - Nature et
Technologie (FRQ-NT). Computations were also performed on the Zwicky cluster at Caltech, supported by the Sherman
Fairchild Foundation and by NSF award PHY-0960291.

\bibliographystyle{iopart-num}
\bibliography{References/References.bib}

\appendix

\section{Adaptive grid structure}
\label{sec:grid}

An important difficulty when simulating NSBH mergers with hot nuclear-theory based equations of state is that, when the neutron star
disrupts close to the ISCO, the fluid forms a very narrow, rapidly moving tidal tail~\cite{Foucart:2014nda}. The difference in length scale between the width of the tail
$\sim 5\,{\rm km}$, the radius of the forming accretion disk $\sim 50\,{\rm km}$, and the distance to which we have to follow
the ejected material to reliably extract its properties $\sim 1000\,{\rm km}$ makes it computationally impossible to study these mergers
without the use of mesh refinement techniques. Previously, we used a fixed mesh refinement method, with a grid composed of nested
boxes of increasing resolution centered on the black hole (with a factor of two change in resolution between each level of refinement). During
the disruption of the neutron star, this is very suboptimal, as the narrow tidal tail has a very low filling factor in the cubic grid, $(10-20)\,\%$, before
formation of an accretion disk. This leads to wasted numerical resources in a phase of the merger during which a high numerical resolution is 
required to avoid numerical shocks in the contracting tidal tail.

To solve this issue, we now let SpEC automatically turn small sections of the computational domain on and off, as matter moves through the grid.
Because SpEC evolves the spacetime metric on a separate grid, there is no need for the finite volume grid to extend to regions where
no matter is present. In practice, during the disruption of the neutron star, we use $7$ levels of refinement, each with $200^3$ 
points\footnote{These $7$ refinement levels form a purely potential grid -- during the simulations, we never need to turn on subdomains in more than
$5$ levels of refinement at any given time.}. Each level is subdivided into $8^3$ subdomains, which can be individually turned on and off. The
central $4^3$ subdomains of each level are always turned off, except at the finest level of refinement. (They overlap subdomains with
a finer mesh.) A subdomain is turned on when, anywhere within $6$ grid spacings of its outer boundary, the fluid density satisfies
\beq
\rho > \rho_{FMR} = 3\times10^{10}\left[0.001+\left(\frac{2r_{\rm exc}}{r+r_{\rm exc}}\right)^2\right]\,{\rm g\,cm^{-3}},
\eeq
with $r_{\rm exc}$ the excision radius and $r$ the coordinate distance to the black hole center in grid coordinates. 
A subdomain is turned off as soon as all grid cells within $6$ grid spacings of its boundary 
satisfy the condition $\rho<(\rho_{\rm FMR}/2)$. We test these conditions every $20$ time steps, and modify the domain (including rebalancing the computational
load) whenever at least one subdomain needs to be turned on. We note that, while this method allows for more efficient evolutions,
it does not match the flexibility of a true Adaptive Mesh Refinement technique. In our algorithm, the structure of the numerical grid
is fixed by the user, and the code can only choose whether to ignore a region of spacetime or not. In essence, this is thus still a Fixed Mesh Refinement
algorithm, but one which is capable of ignoring the large fraction of the numerical grid in which no matter is present in neutron star-black hole mergers.

\begin{figure}
\begin{center}
\includegraphics*[width=0.45\textwidth]{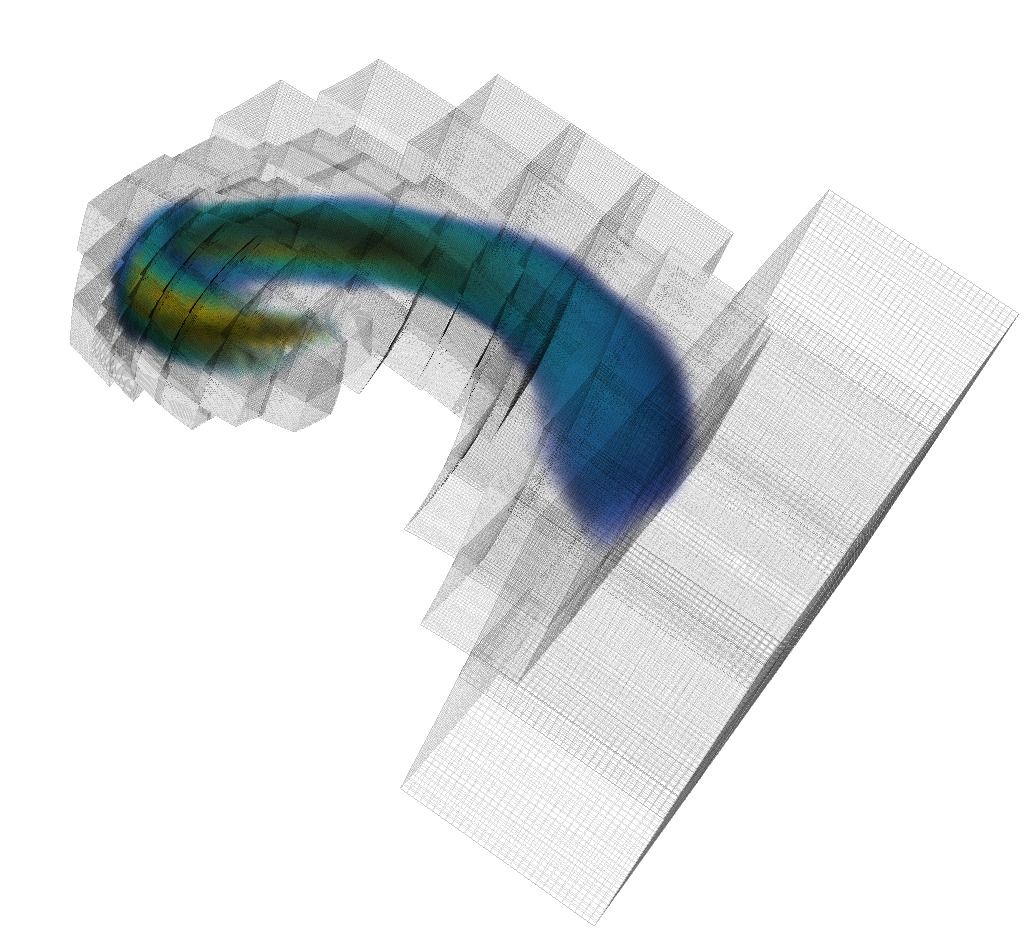}
\includegraphics*[width=0.5\textwidth]{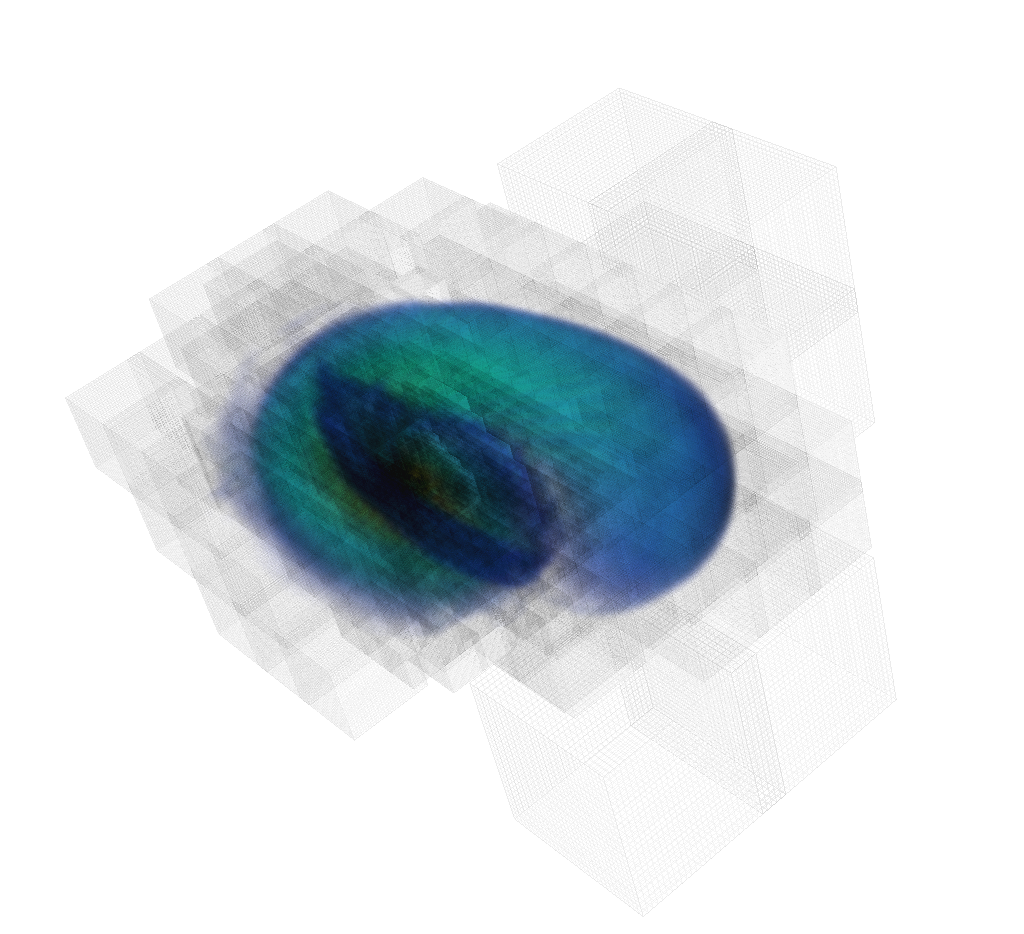}
\caption{Grid structure and baryon density during the disruption of the neutron star ({\it left}), and after disk formation ({\it right}). Both images are snapshots of model
M5-S7-I60, with the color scale showing the logarithm of the baryon density wherever $\rho \gtrsim 10^{10}{\rm g/cm^3}$. The advantage of an adaptive grid is clearly visible during disruption. After merger, the technique remains advantageous
in following the tidal tail, while the inner regions are more densely filled with disk material.}
\label{fig:FMR}
\end{center}
\end{figure}

Examples of the resulting domain decomposition are shown on Fig.~\ref{fig:FMR}. During tidal disruption, our algorithm significantly reduces the cost of the
evolution by only using subdomains covering the fluid. After disk formation, we still gain by only turning on subdomains covering the disk and
tidal tail, but the filling factor of the matter is now significantly larger ($50\%-80\%$ at the finest level of refinement). Once a disk forms, the resolution
requirements to accurately evolve the fluid become less stringent. We remove the finest level of refinement around the time at which the mass accretion
rate onto the black hole reaches its first local minimum (i.e. before circularization of the disk material begins), and the second finest level once a
more axisymmetric disk has been formed.

This adaptive mesh is only used during tidal disruption (or plunge) of the neutron star and for the post-merger evolution of the remnant. Before merger,
we use a uniform grid with spacing $\Delta x = 220\>{\rm m}$ in the coordinates of the finite volume grid. To determine the grid spacing in the laboratory frame,
we have to remember that SpEC uses a time-dependent map
between the grid coordinates and the laboratory frame. A first map keeps the center of the compact objects fixed in the coordinates of the 
pseudospectral grid on which we evolve Einstein's equations, and makes sure that the excised region inside the apparent horizon of the black hole
is of a shape and size guaranteeing that no information flows from the boundary into the computational domain~\cite{Hemberger:2012jz}. 
A second map relates the coordinates of the finite
volume grid to the coordinates of the pseudospectral grid, and keeps the neutron star fluid entirely within the finite volume grid.
The grid spacing in the laboratory frame is thus time-dependent. Before merger, the grid
spacing will change by $\sim 25\%$ around the initial value $\Delta x = 220\>{\rm m}$, with a better resolution in the directions in which tides cause the 
neutron star to contract, and a lower resolution in the direction in which it expands. 

During the tidal disruption of the neutron star and the post-merger evolution, the finite volume grid and pseudospectral grid use the same coordinate system.
The adaptive grid described above is used to keep the neutron star fluid on the finite volume grid. We still use a time-dependent map between the grid
coordinates and the laboratory frame, to control the location, shape, and size of the apparent horizon of the black hole. 
We keep the grid spacing of the finest level of refinement in the grid coordinates to $\Delta x \approx 250\,{\rm m}$. Because
the grid coordinates contracted as the binary inspiralled, this generally corresponds to a finer resolution in the laboratory frame,
which will be different for each simulation. 
We provide the typical value of the finest grid spacing during disruption for simulations at our fiducial resolution in Table~\ref{tab:ID}, and discuss
lower resolution simulations and the inferred numerical accuracy of our results in Sec.~\ref{sec:errors}.

\section{Numerical accuracy}
\label{sec:errors}

Beyond the fiducial simulations using an adaptive grid described in~\ref{sec:grid}, we evolve each model at one or more lower resolution, using
fixed finite volume grids with 2-3 levels of refinement. These lower resolution simulations typically use $\Delta x_{\rm dis}\approx(250-300)\,{\rm m}$ for the grid spacing  
$\Delta x_{\rm dis}$ of the finest refinement level in the laboratory frame, instead of $\Delta x_{\rm dis}\approx(150-200)\,{\rm m}$ with the adaptive grid 
(see Table~\ref{tab:ID}). We use these lower resolution simulations to get estimates of our numerical accuracy.

The main source of error in our simulations is the evolution of the fluid close to the horizon of the black hole, and thus to the surface of the region inside the black hole 
which is excised from our computational domain. For fluid elements 
which plunge into the black hole and are causally disconnected from the rest of the matter, this has no practical effect on the evolution. However, for models in which
the tidal disruption of the neutron star occurs very close to the black hole, this is a significant source of error.
As for higher mass systems tidal disruption occurs closer to the black hole horizon, numerical errors are typically larger. 
This is compounded by the fact that
the disrupted neutron star forms a narrow, hard-to-resolve tidal tail with a sharp front edge moving at supersonic speeds through the grid. 
This difficulty in resolving the disruption of neutron stars described by hot, nuclear-theory based equations of state was already noted for aligned-spin 
binaries~\cite{Foucart:2014nda}. Merger simulations with polytropic and piecewise-polytropic equations of state, on the other end, do not appear to show the 
creation of such narrow tails, making their disruption significantly easier to simulate.

The worst manifestation of this numerical error is the ejection at high velocity of some material which would otherwise fall into the black hole, due to numerical shocks
in the under-resolved tidal tail. This effect is clearly visible in our lower resolution simulations for all simulations with $M_{\rm BH}=7M_\odot$, and is the main motivation
to use an adaptive grid and increase the resolution of our simulations (the evolution of the fluid at distances larger than $\sim 1.5 r_{\rm H}$, with $r_H$ the radius of 
the horizon, is well converged for all of our simulations). This {\it numerical ejecta} can be easily distinguished from the physical {\it dynamical ejecta}, as it originates
in a completely different part of the disrupting neutron star. It clearly converges away at high-resolution, except for model M7-S9-I60. Model M7-S9-I60 is the most
problematic case because it is a relatively high mass ratio, and a case which is only marginally disrupting. The production of a few percents of a solar mass
of unphysical ejecta even at a resolution $\Delta x \approx 100\,{\rm m}$ significantly affects measurements of the ejected mass and, potentially, 
baryon mass outside of the black hole. Accordingly, we do not report post-merger quantities for that model.

For all other disrupting models, the mass of the dynamical ejecta converges to $\lesssim 20\%$, while the baryon mass outside of the black hole and the other
reported properties of the disk and ejecta (average velocities, composition, temperature, inclination) converge to $\lesssim 10\%$. Finally, for the non-disrupting
model M7-S7-I60, we report upper bounds on the ejected mass and baryon mass outside the black hole, obtained from the highest resolution simulation that we 
performed.

\end{document}